# Modeling of Multisite Precipitation Occurrences Using Latent Gaussian-based Multivariate Binary Response Time Series

**Hsien-Wei Chen[a]**

[a] Department of Civil Engineering and Applied Mechanics, McGill University, 817 Sherbrooke Street West, Montreal, Quebec H3A 0C3, Canada

**Abstract**

A new stochastic model for daily precipitation occurrence processes observed at multiple locations is developed. The modeling concept is to use the indicator function and the elliptical shape of multivariate Gaussian distribution to represent the joint probabilities of daily precipitation occurrences. By using this concept, the number of parameters needed for precipitation occurrence modeling can be largely reduced when compared to the commonly used two-state Markov chain approach. With this parameter reduction, the modeling of spatio-temporal dependence of daily precipitation occurrence processes observed at different locations is no longer difficult. Results of an illustrative application using the precipitation record available from a network of ten raingauges in the southern Quebec region, also demonstrate the accuracy and the feasibility of the proposed model.

**Highlight**

- A solution to the spatio-temporal dependence modeling of precipitation occurrences
- A multisite precipitation occurrence model for practical applications
- A multivariate binary response time series model

## 1. Introduction

To study drought, the stochastic modeling of precipitation occurrences is a useful tool for assessing the drought extend and frequency. Especially, for studying extreme events with a low frequency, a stochastic model allows to generate synthesis precipitation occurrence sequences that are statistically similar to historical records. Furthermore, these generated sequences are long enough for making inference on the low frequency extreme events, which are sometimes difficult to examine from historical observations. More importantly, the spatial variation that occurrence models describe is a crucial element for examining the spatial intermittency of precipitation, which should also be taken into consideration for accurate flood estimations.

To ensure that the generated synthesis sequences of precipitation occurrences are statistically resemble to and can therefore represent historical records, certain statistical properties need to be taken into consideration by the stochastic model. For at-site statistical properties, the temporal dependence of precipitation occurrences is probably the main consideration because it represents the tendency of dry/wet day occurrences after a dry/wet day. For drought research, this tendency is especially important since it corresponds to the lengths of consecutive dry days. Therefore, the temporal dependence is a fundamental element that needs to be captured for precipitation occurrence modeling. The stochastic modeling based on this wet-to-dry/dry-to-wet transition tendency can be found in many applications (Harrold et al. 2003; Katz, 1977; Richardson, 1981; Roldan and Woolhiser, 1982; Srikanthan and McMahon, 2001; Todorovic and Woolhiser, 1975). However, this type of precipitation occurrence modeling approach mainly focuses on the occurrence process at a single location. Direct applications of this type of single site models to different raingauge sites may ignore the spatial variation of precipitation occurrence processes. Therefore, for inter-site statistical properties, the spatial dependence of precipitation occurrences should also be taken into consideration. As a consequence, the multisite modeling of occurrence processes is much more complex than the single site modeling. The popular solution is to utilize climate variables as predictors (Gachon et al. 2005) with statistical models that can deal with multiple variables (Fahrmeir and Tutz, 1994; Yakowitz, 1985) to describe the spatial dependence. For the regression-type model, studies can be found in Asong et al. (2016), Chandler and Wheater (2002), and Yang et al. (2005). For the non-parametric-type model, application results are provided in Apipattanavis et al., (2007), Buishand and Brandsma (2001), Mehrotra and Sharma, (2007), and Steinschneider and Brown (2013). However, to capture both the temporal and the spatial dependences is still challenging. Often, either the temporal or the spatial dependence is traded-off for better accuracy of the other. Detailed reviews that illustrate this challenge are provided in Buishand and Brandsma (2001), Frost et al. (2009, 2011), Mehrotra et al. (2006), and Yang et al. (2005).

Nonetheless, Wilks' (1998) illustrated an approach for occurrence modeling that overcomes this challenge. This approach essentially starts with single site modeling of precipitation occurrences, and then introduces inter-station correlations to force these mutually independent single site models to have spatial dependence as the observations. However, this approach has a limitation: the spatial dependence with time lags cannot be captured. In other words, the tendency of dry/wet day occurrences after a dry/wet day is observed at a different raingauge site, cannot be depicted by using this approach. Therefore, for watershed areas with a time of concentration longer than one day, it would be appreciated to have a model that can describe this spatio-temporal dependence in additional to the temporal and the spatial dependences to have an even more accurate estimation on drought and flood.

In viewing the above mentioned issues, this study aims to develop a stochastic model for precipitation occurrences with more flexibility in accurately describing both the extremes of consecutive dry days and the spatio-temporal variation of precipitation occurrences. The feasibility and the potential of the proposed model in practical applications will be shown via the precipitation record available in the southern Quebec region.

## 2. Methodology

The Markov chain model is a commonly used approach to model precipitation occurrence time series (Roldan and Woolhiser, 1982; Srikanthan and McMahon, 2001) since the Markov chain model is built by directly fitting the joint distribution of binary responses to the temporal dependence structure of observed rainfall occurrences. However, approximately $2^{s \times (r+1)}$ parameters are required when building a lag-$r$ Markov chain model for $s$-variable time series. This exponential growth of the number of required parameters limits the application of long-lag Markov chain models to multivariate time series. Therefore, a new model for multivariate binary response time series is needed.

In this study, the elliptical shape of multivariate Gaussian density is used to avoid directly fitting the model to the observed joint probabilities of binary response realizations. Consequently, while preserving the major spatio-temporal statistical properties, the number of the required parameters for modeling is largely reduced. Then, only approximately $C_2^{s \times (r+1)}$ (the number of two-combinations of $s \times (r+1)$ distinct elements) parameters are needed for modeling a lag-$r$, $s$-variable binary response time series. Thus, a simpler multivariate time series model for binary responses can be developed for modeling daily precipitation occurrences at different sites.

*Model description.* Assume that the random variable of the binary response $O$ is the result of the indicator function acting on the standard Gaussian random variable $Z$ with the constant threshold $C$ and can be expressed as the following.

$$O_{t,i} = 1(Z_{t,i} > C_i) \ \text{ for } \ i = 1, 2, \cdots s \ \text{ and } \ t = 1, 2, \cdots \quad (1)$$

where the subscript $t$ represents the time, the subscript $i$ indicates the belonging time series, and $1(\cdot)$ represents the indicator function that equals to one when the statement inside the parenthesis is true and zero when is false. Then, the probability of $O$ equals to one is a function of $C$. In addition, the assumption that $C$ does not change with time ensures that this probability remains the same with time; i.e. $p_i = P(O_{t,i} = 1)$ is a constant.

Furthermore, the following assumptions are made to ensure that the correlation structure of the binary response time series do not change with time.

$$\Sigma_k = Cov(\tilde{Z}_t, \tilde{Z}_{t-k}) = E\left[(Z_{t,1}, Z_{t,2}, \cdots, Z_{t,s})^T (Z_{t-k,1}, Z_{t-k,2}, \cdots, Z_{t-k,s})\right] = const. \, matrix$$

(2)

$$\text{for} \ \ k = 0, 1, 2, \cdots$$

*Parameter estimation.* As shown by Equations (1) and (2), $\tilde{C} = (C_1, C_2, \cdots, C_s)^T$ and $\Sigma_k$ ($k = 0, 1, \cdots, r$) need to be estimated for calibrating a lag-$r$, $s$-variable model. Furthermore, since $C_i$ is directly related to the marginal probability $p_i$, it is natural to use the following consistent estimator of $C_i$.

$$\hat{C}_i = \Phi^{-1}(1 - \hat{p}_i) \ \ \text{with} \ \ \hat{p}_i = \frac{1}{n_i}\sum_t O_{t,i}$$

(3)

where $\Phi$ is the cumulative distribution function (CDF) of the standard normal distribution, $n_i$ is the number of observed $O_{t,i}$, and $\hat{C}_i$ and $\hat{p}_i$ represent the estimators of the parameters $C_i$ and $p_i$, respectively.

Finally, by using the estimated $\tilde{C}$ obtained from Equation (3), the consistent estimator of $\Sigma_k$ can be obtained by solving the following equation for the off-diagnoal elements of $\Sigma_k$.

$$\Phi_2(\hat{C}_u, \hat{C}_v \mid \{\Sigma_k\}_{u,v}) = 1 - \hat{p}_u - \hat{p}_v + \hat{p}_{u,v}^k \ \ \text{with} \ \ \hat{p}_{u,v}^k = \frac{1}{n_{u,v}^k}\sum_t \left[O_{t,u} \times O_{t-k,v}\right]$$

(4)

where $\Phi_2(\cdot, \cdot \mid \rho)$ is the CDF that describes the joint probability distribution of a pair of standard Gaussian random variables, $\rho$ represents the correlation coefficient between these two standard Gaussian random variables, $n_{u,v}^k$ is the number of observed $O_{t,u} \times O_{t-k,v}$ pairs, and $\{\Sigma_k\}_{u,v}$ represents the $u$ row, $v$ column element of $\Sigma_k$. Also, it is not difficult to solve Equation (4) numerically since highly efficient methods to calculate the bivariate Gaussian CDF have been developed (Genz, 2004).

To illustrate the concept, consider a pair of binary response random variables, $O_1$ and $O_2$, and the corresponding $Z_1$, $Z_2$, $C_1$, and $C_2$. As shown in Figure 1, the threshold $C_1$ and $C_2$ (red lines) slice the bivariate Gaussian density of $(Z_1, Z_2)^T$ into four areas: Ⅰ, Ⅱ, Ⅲ, and Ⅳ. As a result, observations of $(Z_1, Z_2)^T$ falling on area Ⅰ, Ⅱ, Ⅲ, and Ⅳ represent the occurrence of $\{O_1 = 1, O_2 = 1\}$, $\{O_1 = 0, O_2 = 1\}$, $\{O_1 = 0, O_2 = 0\}$, and $\{O_1 = 1, O_2 = 0\}$ respectively. Therefore, Equation (3) is essentially looking for the values of $C_1$ and $C_2$ that make the probability that $(Z_1, Z_2)^T$ falls on area Ⅰ or Ⅳ, i.e. $P(O_1 = 1)$, equal to the observed probability of $P(O_1 = 1)$, and the probability that $(Z_1, Z_2)^T$ falls on area Ⅰ or Ⅱ, i.e. $P(O_2 = 1)$, equal to the observed probability of $P(O_2 = 1)$.

Similarly, Equation (4) is looking for the value of the correlation coefficient between the two standard Gaussian random variables $Z_1$ and $Z_2$ illustrated in Figure 1 that makes the probabilities that $(Z_1, Z_2)^T$ falls on area Ⅰ and Ⅲ, i.e. $P(O_1 = 1, O_2 = 1)$ and $P(O_1 = 0, O_2 = 0)$, equal to the observed probabilities of $P(O_1 = 1, O_2 = 1)$ and $P(O_1 = 0, O_2 = 0)$, respectively, when $C_1$ and $C_2$ are determined by Equation (3). When $Corr(Z_1, Z_2)$ is positive, the long axis of the ellipses of the contour falls on area Ⅰ and Ⅲ. On the contrary, when $Corr(Z_1, Z_2)$ is negative, the long axis of the ellipses of the contour falls on area Ⅱ and Ⅳ. Therefore, it is clear that $P(O_1 = 1, O_2 = 1)$ and $P(O_1 = 0, O_2 = 0)$ monotonically increase/decrease when $Corr(Z_1, Z_2)$ increases/decreases. Furthermore, when $Corr(Z_1, Z_2)$ is equal to one, $P(O_1 = 1, O_2 = 1)$ and $P(O_1 = 0, O_2 = 0)$ are equal to $\min\{P(O_1 = 1), P(O_2 = 1)\}$ and $\min\{P(O_1 = 0), P(O_2 = 0)\}$, respectively, and when $Corr(Z_1, Z_2)$ is equal to negative one, one of $P(O_1 = 1, O_2 = 1)$ and $P(O_1 = 0, O_2 = 0)$ is zero. Based on the above, the solution to Equation (4) always exists and is unique since a joint probability is always less than or equal to its corresponding marginal probabilities and greater than or equal to zero.

Equation (3) and Equation (4) together ensure that the bivariate model of the binary response pair $O_1$ and $O_2$ has exactly the same joint probability as the observed joint probability. If an additional binary response $O_3$ is introduced, Figure 1 will become a three-dimensional plot. $C_1$, $C_2$, and $C_3$ together slice the three-dimensional space into eight areas. Instead of directly fitting these eight areas to their corresponding observed moments, Equation (3) and Equation (4) provide a modeling approach by looking at the two additional two-dimensional projections of the three-dimensional plot, i.e. $Z_1$ versus $Z_3$ and $Z_2$ versus $Z_3$. This modeling approach not only largely reduces the number of parameters of the multivariate model but also does not lose too much information since the joint probabilities of more than two binary response random variables are smaller in value than the joint probabilities of a pair of these binary response random variables (exponentially decay with the number of binary response random variables) and difficult to accurately estimate from observations.

*Simulation*. To simulate the multivariate binary response time series is straightforward since it is essentially simulating multivariate Gaussian time series. Once $\Sigma_k$ ( $k = 0, 1, \cdots, r$ ) is estimated, the simulation of the time series $\widetilde{Z}_t$ can be achieved by Equations (5) and (6).

$$
\begin{pmatrix} \widetilde{Z}_{r+1} \\ \vdots \\ \widetilde{Z}_1 \end{pmatrix} \sim N_{s(r+1)}\left(\vec{0}, \Sigma_{all}\right) \text{ with } \Sigma_{all} = \left[\begin{array}{c|c} \Sigma_0 & \Sigma_{all(1,2)} \\ \hline \Sigma_{all(2,1)} & \Sigma_{all(2,2)} \end{array}\right] = \begin{bmatrix} \Sigma_0 & \Sigma_1 & \Sigma_2 & \cdots & \Sigma_r \\ \Sigma_1^{\ T} & \Sigma_0 & \Sigma_1 & \cdots & \Sigma_{r-1} \\ \vdots & \ddots & \ddots & \ddots & \vdots \\ \Sigma_{r-1}^{\ \ T} & \cdots & \Sigma_1^{\ T} & \Sigma_0 & \Sigma_1 \\ \Sigma_r^{\ T} & \cdots & \Sigma_2^{\ T} & \Sigma_1^{\ T} & \Sigma_0 \end{bmatrix} \quad (5)
$$

$$
\widetilde{Z}_t \mid \widetilde{Z}_{t-1}, \ldots \widetilde{Z}_{t-r} \sim N_s\left( \Sigma_{all(1,2)} \Sigma_{all(2,2)}^{\ \ -1} \left( \widetilde{Z}_{t-1}^{\ \ T}, \cdots, \widetilde{Z}_{t-r}^{\ \ T} \right)^T, \Sigma_0 - \Sigma_{all(1,2)} \Sigma_{all(2,2)}^{\ \ -1} \Sigma_{all(2,1)} \right) \quad (6)
$$

The application of Equation (5) is to simulate unconditional realizations for the first $r+1$ time steps, and the application of Equation (6) is to simulate realizations that are conditional on previously simulated $r$ time steps. Then, the simulated realizations of the multivariate binary time series $\widetilde{O}_t = (O_{t,1}, O_{t,2}, \cdots, O_{t,s})^T$ can be obtained by applying Equation (1) to the estimated $\widetilde{C}$ and the simulated realizations of the multivariate Gaussian time series $\widetilde{Z}_t$.

However, when the estimated covariance matrix $\sum_{all}$ is nearly singular, the simulation of $\tilde{Z}_t$ is difficult. As a result, Equations (7) and (8) are adopted to adjust the estimated covariance matrix $\sum_{all}$.

$$\sum_{all} + \varepsilon_1(1 - I_{s(r+1)}) = P\Lambda P^T \tag{7}$$

$$\sum_{all}{}^{adj.} = \Omega P\Lambda^{adj.}P^T\Omega \tag{8}$$

where $I_{s(r+1)}$ is a $s(r+1)$ by $s(r+1)$ identity matrix, $\varepsilon_1$ is a constant, $P$ is the collection of the eigenvectors of $\sum_{all} + \varepsilon_1(1 - I_{s(r+1)})$, and $\Lambda$ is a diagonal matrix in which the $i^{th}$ diagonal element is the corresponding eigenvalue of the $i^{th}$ column eigenvector of $P$. $\Lambda^{adj.}$ is the adjusted $\Lambda$ in which the diagonal elements less than the constant $\varepsilon_2$ are forced to be $\varepsilon_2$. $\Omega$ is a diagonal matrix in which the diagonal elements are the inverse of the square root of the corresponding diagonal elements of $P\Lambda^{adj.}P^T$. Therefore, the adjusted covariance matrix $\sum_{all}{}^{adj.}$ is obtained by first manually selecting an appropriate small value for $\varepsilon_2$ (the smallest eigenvalue of $\sum_{all}$ is a good reference) then finding the optimal value for $\varepsilon_1$ that makes the average of all elements within $\sum_{all}{}^{adj.} - \sum_{all}$ zero.

The adjustment of $\Lambda$ is to avoid the extremely small value of the determinant of $\sum_{all}$, which equals to the product of its corresponding eigenvalues. However, the adjustment of $\Lambda$ also increases the diagonal elements of $\sum_{all}$, which are the variance of the standard normal distribution and are supposed to be one. As a result, $\Omega$ is applied to force them to equal one. However, the application of $\Omega$ decreases the values of all the off-diagonal elements of $\sum_{all}$, which are the estimated correlation coefficients. Therefore, the constant $\varepsilon_1$ is introduced to minimize the difference between $\sum_{all}$ and $\sum_{all}{}^{adj.}$.

The concern of this adjustment is that $\sum_{all}^{adj.}$ is different from $\sum_{all}$, and, as a result, the multivariate time series model is affected. However, there are only negligible changes on each value of elements within $\sum_{all}$ after the application of Equations (7) and (8). As shown in Figure 2, this adjustment can correct the singularity of the covariance matrix while it keeps each estimated correlation within $\sum_{all}$ nearly unchanged. Therefore, this modification will not affect the multivariate binary time series model as a whole.

As illustrated in Figure 3, to implement the proposed method for modeling and simulating precipitation occurrences, the precipitation observations collected from raingauge sites needed to be converted to the corresponding multivariate binary response time series by assigning respectively the values zero and one to the dry and wet periods in the time series of each site. With these binary response time series converted from the precipitation observations, Equation (3) can be adopted to estimate the vector $\tilde{C}$. Then, each off-diagonal element of the covariance matrix $\sum_{all}$ can be estimated via Equations (4) and (5). When this estimated covariance matrix is nearly singular, Equations (7) and (8) need to be applied for the adjustment before simulation. To simulate the corresponding binary response time series that represent the precipitation occurrences of the raingauge sites, this adjusted/estimated covariance matrix needs to be adopted to simulate the multivariate Gaussian time series described in Equations (5) and (6). Finally, the simulated realizations for the precipitation occurrences of the raingauge sites can be obtained by applying Equation (1) with the estimated $\tilde{C}$ to the simulated multivariate Gaussian time series.

## 3. Study Area

In this study, the daily precipitation observations of southern Quebec were used to access the feasibility of the modeling approach. The spatial distribution of the raingauge sites is shown in Figure 4, and the altitude of each raingauge site is listed in Table 1.

|  | Latitude | Longitude | Height (m) |
| --- | --- | --- | --- |
| Dorval | 45.4667 N | 73.7500 W | 36.00 |
| Cornwall | 45.0156 N | 74.7489 W | 64.00 |
| Drummondville | 45.8833 N | 72.4833 W | 82.30 |
| Farnham | 45.3000 N | 72.9000 W | 68.00 |
| Lennoxville | 45.3689 N | 71.8236 W | 181.00 |
| Morrisburg | 44.9236 N | 75.1883 W | 81.70 |
| Oka | 45.5000 N | 74.0667 W | 91.40 |
| Ottawa CDA | 45.3833 N | 75.7167 W | 79.20 |
| St. Alban | 46.7167 N | 72.0833 W | 76.20 |
| St. Jerome | 45.8000 N | 74.0500 W | 169.50 |

Table 1. The location and altitude of the raingauge sites.

The temporal coverage of the daily precipitation records of all raingauge sites ranges uniformly from 1961-Jan-01 to 2001-Dec-31. Furthermore, the data set is separated into two groups for the calibration (1961-Jan-01 to 1985-Dec-31) and validation (1986-Jan-01 to 2001-Dec-31) purposes, respectively. Table 2 and Table 3 show the quality of the data set by indicating the number of missing observations, which are unavoidable for precipitation records.

| | Jan | Feb | Mar | Apr | May | Jun | Jul | Aug | Sep | Oct | Nov | Dec |
|---|---|---|---|---|---|---|---|---|---|---|---|---|
| Dorval | 0 | 0 | 0 | 0 | 0 | 0 | 0 | 0 | 0 | 0 | 0 | 0 |
| Cornwall | 48 | 14 | 68 | 40 | 40 | 38 | 8 | 7 | 3 | 16 | 14 | 50 |
| Drummondville | 5 | 41 | 17 | 7 | 45 | 38 | 80 | 71 | 37 | 36 | 4 | 40 |
| Farnham | 0 | 1 | 2 | 0 | 2 | 0 | 0 | 4 | 2 | 1 | 2 | 6 |
| Lennoxville | 2 | 1 | 2 | 0 | 1 | 3 | 1 | 8 | 0 | 2 | 5 | 3 |
| Morrisburg | 5 | 2 | 2 | 0 | 0 | 0 | 0 | 0 | 5 | 0 | 7 | 2 |
| Oka | 6 | 13 | 4 | 2 | 0 | 4 | 45 | 1 | 0 | 33 | 8 | 6 |
| Ottawa CDA | 0 | 1 | 0 | 0 | 0 | 0 | 0 | 0 | 0 | 0 | 0 | 0 |
| St. Alban | 9 | 7 | 35 | 0 | 1 | 1 | 0 | 19 | 1 | 0 | 3 | 11 |
| St. Jerome | 6 | 11 | 6 | 32 | 33 | 74 | 40 | 2 | 7 | 2 | 3 | 11 |
| Total number of days | 775 | 706 | 775 | 750 | 775 | 750 | 775 | 775 | 750 | 775 | 750 | 775 |

Table 2. The number of missing observations in the calibration period (1961-Jan-01 to 1985-Dec-31). Except the observations with trace flags are counted as dry days, all the observations with flags are also treated as missing observations.

| | Jan | Feb | Mar | Apr | May | Jun | Jul | Aug | Sep | Oct | Nov | Dec |
|---|---|---|---|---|---|---|---|---|---|---|---|---|
| Dorval | 0 | 0 | 0 | 1 | 0 | 0 | 0 | 0 | 0 | 0 | 0 | 0 |
| Cornwall | 9 | 8 | 3 | 2 | 5 | 4 | 2 | 0 | 0 | 5 | 4 | 5 |
| Drummondville | 4 | 1 | 6 | 31 | 31 | 13 | 0 | 1 | 1 | 5 | 3 | 6 |
| Farnham | 31 | 31 | 32 | 1 | 1 | 1 | 1 | 67 | 61 | 62 | 68 | 31 |
| Lennoxville | 36 | 28 | 8 | 21 | 5 | 12 | 5 | 46 | 13 | 35 | 49 | 44 |
| Morrisburg | 3 | 4 | 0 | 3 | 2 | 2 | 0 | 1 | 1 | 1 | 30 | 0 |
| Oka | 8 | 2 | 4 | 0 | 0 | 0 | 0 | 30 | 11 | 0 | 0 | 9 |
| Ottawa CDA | 0 | 0 | 1 | 0 | 0 | 0 | 0 | 0 | 0 | 0 | 0 | 0 |
| St. Alban | 8 | 1 | 4 | 0 | 0 | 5 | 22 | 1 | 0 | 0 | 9 | 4 |
| St. Jerome | 5 | 8 | 39 | 5 | 2 | 0 | 2 | 1 | 1 | 0 | 22 | 31 |
| Total number of days | 496 | 452 | 496 | 480 | 496 | 480 | 496 | 496 | 480 | 496 | 480 | 496 |

Table 3. The number of missing observations in the validation period (1986-Jan-01 to 2001-Dec-31). Except the observations with trace flags are counted as dry days, all the observations with flags are also treated as missing observations.

## 4. Results and Discussion

In this study, a day is considered a wet day ($O = 1$) when the accumulated precipitation of that day is greater than or equal to $1\,mm$; otherwise, it is considered a dry day ($O = 0$). Then, the latent Gaussian-based multivariate binary response time series modeling approach with the consideration of lag-2 temporal dependences is applied to fit the observed daily precipitation occurrences in the calibration period, i.e. $r = 2$ and $s = 10$. Furthermore, to account for the seasonality, the daily precipitation occurrences of each calendar month are modeled separately. Finally, one thousand sets of simulations of both the calibration and validation periods are produced and compared with the observed daily precipitation occurrences to evaluate the model's performances. For each set, the daily precipitation occurrence records in both the calibration and the validation period are generated by the proposed model. The performances of the model are evaluated by determining whether the simulated daily precipitation occurrences reproduce similar statistical properties of indices of practical interest as the observed daily precipitation occurrences produce (Gachon et al., 2005; STARDEX, 2005; Wilks, 1998). These indices include the percentage of wet days, lagged interstation correlations, and maximum number of consecutive dry days, which represent the characteristics of the mean, the spatio-temporal dependence, and the extremes of daily precipitation occurrences, respectively. Moreover, as a comparison to the latent Gaussian model, Wilks' (1998) approach is also applied to the same study area with the same threshold of wet and dry days and exactly the same calibration and simulation settings.

*Percentage of wet days.* As indicated in the methodology section, the latent Gaussian model could fit well the probability of the daily precipitation occurrence at each raingauge site to the observation. The simulated daily precipitation occurrences have exactly the same mean percentage of wet days for each month and for each station as the daily precipitation occurrences in the calibration period. As a result, the evaluation of the percentage of wet days of each single site is of less interest. Consequently, for the percentage of wet days at the seasonal scale, a more practical performance evaluation approach is to compare the empirical distributions of the simulated and observed percentages of wet days collected from each raingauge station in each season of each year. As shown in Figure 5, the latent Gaussian model and Wilks' approach give similar simulation results for the seasonal percentage of wet days. Both models reasonably fit the observed distribution of the seasonal percentage of wet days in both calibration and validation periods for all four seasons. The most significant bias among the simulation results of both models can be found in the fall season of the validation period. However, when taking the sampling uncertainties of stochastic models into consideration, this bias cannot be viewed as a failure of both models. Both the latent Gaussian model and Wilks' approach can be considered as successful stochastic models in describing the statistical properties of the percentage of wet days at the seasonal scale.

*Lagged interstation correlations.* For the model performances in describing the spatio-temporal dependence, lagged interstation correlations are adopted as evaluation indices. However, because the uncertainty of lagged interstation correlations is of less interest, only one set of lagged interstation correlations is calculated from all the station pairs in each calendar month for both the calibration and the validation periods. As a result, scatter plots are adopted for the comparison. Figure 6 shows the comparison between the observed and the simulated (lag-0) interstation correlations of the daily precipitation occurrences, i.e. $Cor(O_{t,a}, O_{t,b})$ for all station pairs $a$ and $b$ ($a \neq b$). As shown in Figure 6, for both the latent Gaussian model and Wilks' approach, the simulated interstation correlations nearly perfectly fit the observed interstation correlations in the calibration period. These nearly perfect fits in the calibration period are simply the consequence of the direct fitting of both models to the spatial dependence of the observed daily precipitation occurrences in the calibration period. For the latent Gaussian model, a small bias can be found at high correlations in the calibration period. This small bias is the consequence of the application of Equations (7) and (8) in adjusting $\Sigma_{all}$ for the simulation purpose. As also shown in Figure 2, the application of Equations (7) and (8) tends to lower the value of high correlations and raise the value of low correlations. However, this small bias shown in Figure 6 is still negligible. In addition, as illustrated in Figure 6, both the latent Gaussian model and Wilks' model can reasonably describe the interstation dependence in the validation period with similar performances. Therefore, the success of the latent Gaussian approach in modeling the interstation dependence of daily precipitation occurrences can be concluded.

Figure 7 evaluates the capability of the model in capturing the lag-1 interstation correlation of the daily precipitation occurrences, i.e. $Cor(O_{t,a}, O_{t-1,b})$ for all station pairs $a$ and $b$ (when $a=b$, the lag-1 autocorrelation of the corresponding site is evaluated). Similar to the lag-0 case, the simulated lag-1 interstation correlations of the latent Gaussian model nearly perfectly fit the observed lag-1 interstation correlations in the calibration period as a result of the characteristics of the modeling approach of the latent Gaussian model. As also shown in Figure 7, the latent Gaussian model is able to reasonably describe the lag-1 interstation dependence of the observed daily precipitation occurrences in the validation period. However, because of the lack of consideration of the lag-1 interstation dependence, the lag-1 interstation correlations simulated by Wilks' approach are randomly distributed on the scatter plots of Figure 7.

Figure 8 compares both the observed and the simulated lag-2 interstation correlations of daily precipitation occurrences, i.e. $Cor(O_{t,a}, O_{t-2,b})$ for all station pairs $a$ and $b$ (when $a=b$, the lag-2 autocorrelation of the corresponding site is evaluated). Due to the characteristics of the modeling approach of the latent Gaussian model, an almost complete fit of the simulated correlations to the observed correlations is again found in the calibration period. Although the lag-2 interstation correlation values for this study are small, the proposed model is better than Wilks' model. As also shown on the right-hand side of Figure 8, Wilks' model, which, on the contrary, does not take the lag-2 interstation dependence into consideration, produces randomly distributed lag-2 interstation correlations on the scatter plots. Because the observed lag-2 interstation correlations are close to zero, the application of a lag-2 model to the daily precipitation occurrences of the study area may not be necessary. However, this result could indicate the capability of the proposed model to describe the lag-2 interstation correlation structure of precipitations. Furthermore, it is possible to have underestimated maximum numbers of consecutive dry days when failing to consider the lag-2 temporal dependence of daily precipitation occurrences (Mehrotra et al., 2006; Woolhiser, 2008). Therefore, the evaluation of the performance of the models in capturing the maximum number of consecutive dry days could be of greater interest.

*Maximum number of consecutive dry days*. The distribution-comparison approach is adopted again to compare the statistical properties of the observed and simulated maximum numbers of consecutive dry days, and Figure 9 shows these comparison results at the seasonal scale. The consecutive dry days across two seasons are not taken into consideration since it is difficult to determine to which season they belong. However, since the calculation of consecutive dry days from both the observation and simulation follows the same criterion, the discussion of the treatment of the consecutive dry days across two seasons is less important. As shown in Figure 9, both the latent Gaussian model and Wilks' model give nearly the same simulation results of the seasonal maximum number of consecutive dry days. The simulated maximum numbers of consecutive dry days of both models closely fit the observed distribution of the seasonal maximum number of consecutive dry days in both calibration and validation periods and all four seasons. Although biases can be found at large seasonal maximum numbers of consecutive dry days, the corresponding cumulative probabilities of these seasonal maximum numbers of consecutive dry days are too close to one to take these biases as a failure of both models. Furthermore, the slightly worse fitting results in the validation period than in the calibration period can again be viewed as sampling uncertainties of stochastic models. Therefore, the success of the latent Gaussian model in describing the statistical properties of the maximum number of consecutive dry days can be concluded.

*Aggregated scales.* To further investigate the appropriateness of the proposed model in describing the precipitation occurrences at larger scales, the mean, the standard deviation, and the interstation correlation of the total number of wet days at aggregated scales are calculated for both the observed precipitation occurrence records and the corresponding simulated precipitation occurrence records in both the calibration and the validation periods. As illustrated in Figures 10, 11, and 12 for the evaluation results at the monthly, the seasonal, and the annual scales, respectively, the proposed model can reasonably describe the mean, the variation, and the cross-site dependence of the total number of wet days at these three aggregated scales despite the significant uncertainty caused by the not-many-enough years of the observed precipitation record.

## 5. Conclusions

In this study, a new stochastic procedure based on the latent Gaussian-based multivariate binary response time series model was developed for modeling the daily precipitation occurrences at different sites. By using the indicator function and the elliptical shape of the multivariate Gaussian distribution, the number of parameters needed to model the joint distribution of binary response random variables can be largely reduced. Then, a simpler multivariate binary response time series model can be obtained. Consequently, the latent Gaussian-based multivariate binary response time series is theoretically able to model multivariate binary response time series with any number of variables and any lag dependence. This flexibility of the proposed model in describing the dependences provides a framework that is general enough for future extensions. When the assumption/information on the long-term trend of precipitation is available via exogenous predictor variables (Mehrotra et al. 2004), this trend assumption/information could be added to the corresponding parameters of the proposed model (e.g. Richardson, 1981; Roldan and Woolhiser, 1982) to further extend the proposed model to the simulation of future scenarios under different climate change assumptions. The incorporation of the non-stationary assumptions in the proposed modeling framework would be a topic of great interest in the near future (e.g. Fatichi et al. 2011; Paschalis et al. 2013). It is also expect to see that the proposed approach inspires researchers in pushing the boundary of knowledge in hydrology research in the context of climate changes by using this proposed stationary model as the base.

The proposed method preserves the dependences between random variables by directly fitting the model parameters to the observed joint probabilities of binary random variables. This approach is similar to the strategy that the Markov chain approaches adopt in describing the dependences. The difference between the two methods is that the Markov chain methods fit, the four conditional probabilities that one random variable is equal to one/zero (wet/dry) given that the other random variable is equal to one/zero (wet/dry), to the observations. The proposed method fits the four joint probabilities of the combinations: both random variables are equal to one, one random variable is equal to zero and the other random variable is equal to one, and both random variables are equal to zero, to the observations. Furthermore, the proposed approach utilizes the elliptical shape of the probability distribution density of the multivariate Gaussian random variables to reduce the number of parameters used when calibrating the model.

The copula approaches are also a popular method to model joint probabilities (e.g. Li and Babovic, 2019a, 2019b). Although Wilks (1998) did not explicitly mention copulas in the original publication, the concept to convert the precipitation occurrences into random variables that follow the uniform distribution adopted in the publication is in fact similar to the modeling strategy that the copula approaches adopt. However, because precipitation occurrences are binary responses (wet or dry) represented by binary response random variables, the direct application of copulas without any adjustment to modeling these discrete random variables of precipitation occurrences for comparison and contrast with the proposed model may be difficult. Nonetheless, it would be of great interest to have this comparison in future research.

To access the feasibility of the proposed model, an illustrative application was carried out using daily precipitation data available in the southern Quebec region. At the seasonal scale, the proposed latent Gaussian model was able to describe the statistical properties of the observed percentage of wet days and maximum number of consecutive dry days of the study area. At the monthly scale, the latent Gaussian model was able to capture the lag-0, lag-1, and lag-2 interstation correlations of the observed daily precipitation occurrences of the study area. Therefore, it can be concluded that the proposed latent Gaussian-based multivariate binary response time series model can be used to describe accurately the basic statistical properties of the daily precipitation occurrence processes at different sites concurrently.

## Acknowledgements

I appreciate the financial support of McGill Engineering Doctoral Awards (MEDA) provided by the Faculty of Engineering of McGill University. Also, I'm grateful for the funding of the National Sciences and Engineering Research Council of Canada (NSERC) via the Research Assistantship awarded. The support of each party was integral during the production of this manuscript.

generalized linear models. *Water Resources Research* **2005**, *41*, W11415.

**Appendix**

To facilitate the readers to implement the proposed logic, sample MATLAB scripts that execute the precipitation occurrence modeling and simulation are provided in this section as the reference. The input matrix: PRECPDATAcal stores the cleaned precipitation record used for the model calibration. The first four columns of the input matrix orderly store the serial date number, year, month, and day with the cleaned precipitation time series of each raingauge site appended from the fifth column. In the time series, positive values represent the precipitation amounts of wet days, the value zero indicates dry days, and –Inf expresses missing observations. The script A2 estimates the vector $\tilde{C}$. The script A3 calculates the covariance matrix $\sum_{all}$. The script A4 adjusts this estimated covariance matrix to resolve the singularity problem via the application of the function indicated by the script attached at the end of the appendix. Finally, the script A5 simulates precipitation occurrences.

```matlab
1  %%%% A2-Estimation of PVEC and CVEC %%%%
2  clear; clc; close all;
3  load('PRECPDATAcal.mat');
4  PVEC=NaN*ones(10,12);
5  CVEC=NaN*ones(10,12);
6  for MONi=1:12;
7      for SITEi=1:10;
8          PVEC(SITEi,MONi)=sum(PRECPDATAcal(PRECPDATAcal(:,3)==MONi,SITEi+4)>0)/sum(PRECPDATAcal↵
(PRECPDATAcal(:,3)==MONi,SITEi+4)>=0);
9          CVEC(SITEi,MONi)=icdf('norm',1-PVEC(SITEi,MONi),0,1);
10     end;
11 end;
12 save('PA2-PVEC','PVEC')
13 save('PA2-CVEC','CVEC')
14
15
16
17
```

```matlab
1  %%%% A3-Estimation of SIGMAkWET %%%%
2  clear; clc; close all;
3  load('PRECPDATAca1.mat');
4  load('PA2-CVEC.mat');
5  load('PA2-PVEC.mat');
6
7  SIGMA0WET=cell(12,1); lagk=0;
8  for lagk=0:2;
9      eval(['SIGMA',num2str(lagk),'WET=cell(12,1);']);
10     for MONi=1:12;
11         TEMPSIGMAmatrix=NaN*ones(10,10);
12         for SITEi=1:10;
13             for SITEj=1:10;
14                 if( (SITEi==SITEj)&&(lagk==0) );
15                     TEMPSIGMAmatrix(SITEi,SITEj)=1;
16                 else
17                     TINDEX=find(PRECPDATAca1(:,3)==MONi); TINDEX=TINDEX(TINDEX>lagk);
18                     OBSERVATION=[PRECPDATAca1(TINDEX,SITEi+4),PRECPDATAca1(TINDEX-lagk,SITEj+4)];
19                     OBSERVATIONWET=OBSERVATION;
20                     OBSERVATIONWET(OBSERVATION>0)=1;
21                     OBSERVATIONWET(OBSERVATION==0)=0;
22                     OBSERVATIONWET(OBSERVATION<0)=NaN;
23                     OBSJOINTWET=prod(OBSERVATIONWET,2);
24                     OBSJOINTPROB=sum(OBSJOINTWET==1)/sum(((OBSJOINTWET==1)|(OBSJOINTWET==0)));
25
26                     TEMPFUN=@(THO) mvncdf([CVEC(SITEi,MONi);CVEC(SITEj,MONi)],[0;0],[1,THO;THO,↵
       1])-(1-PVEC(SITEi,MONi)-PVEC(SITEj,MONi)+OBSJOINTPROB);
27                     TEMPSIGMAmatrix(SITEi,SITEj)=fzero(TEMPFUN,[-1+10^-16,1-10^-16]);
28                 end;
29             end;
30         end;
31         eval(['SIGMA',num2str(lagk),'WET{MONi,1}=TEMPSIGMAmatrix;']);
32     end
33     eval(['save(''PA3-SIGMA',num2str(lagk),'WET(RAW ESTIMATION).mat'',''SIGMA',num2str↵
   (lagk),'WET'');'])
34 end;
35
36
```

```matlab
1  %%%% A4-SIGMAkWET Correction %%%%
2  clear; clc; close all;
3  load('PA3-SIGMA0WET(RAW ESTIMATION).mat');
4  load('PA3-SIGMA1WET(RAW ESTIMATION).mat');
5  load('PA3-SIGMA2WET(RAW ESTIMATION).mat');
6
7  RAWSIGMA0WET=SIGMA0WET; SIGMA0WET=cell(12,1);
8  RAWSIGMA1WET=SIGMA1WET; SIGMA1WET=cell(12,1);
9  RAWSIGMA2WET=SIGMA2WET; SIGMA2WET=cell(12,1);
10
11 ERROR2=0.05;
12 for MONi=1:12;
13     RAWSIGMAa11=[RAWSIGMA0WET{MONi,1}   ,RAWSIGMA1WET{MONi,1}   , RAWSIGMA2WET{MONi,1};
14                 RAWSIGMA1WET{MONi,1}'  ,RAWSIGMA0WET{MONi,1}   , RAWSIGMA1WET{MONi,1};
15                 RAWSIGMA2WET{MONi,1}'  ,RAWSIGMA1WET{MONi,1}'  , RAWSIGMA0WET{MONi,1}];
16     TEMPFUN=@(ERROR1) sum(sum(EIGCORRECTION(RAWSIGMAa11,ERROR1,ERROR2)-RAWSIGMAa11));
17     ADJSIGMAa11=EIGCORRECTION(RAWSIGMAa11,fzero(TEMPFUN,0),ERROR2);
18     SIGMA0WET{MONi,1}=ADJSIGMAa11(1:10, 1:10);
19     SIGMA1WET{MONi,1}=ADJSIGMAa11(1:10,11:20);
20     SIGMA2WET{MONi,1}=ADJSIGMAa11(1:10,21:30);
21 end;
22 save('PA4-SIGMA0WET(ADJ ESTIMATION).mat','SIGMA0WET');
23 save('PA4-SIGMA1WET(ADJ ESTIMATION).mat','SIGMA1WET');
24 save('PA4-SIGMA2WET(ADJ ESTIMATION).mat','SIGMA2WET');
25
26
27 figure; hold on; box on; grid on; xlim([-0.3,1]); ylim([-0.3 1])
28 for MONi=1:12;
29     scatter(reshape(SIGMA0WET{MONi,1},[],1),reshape(RAWSIGMA0WET{MONi,1},[],1),21,'r.')
30     scatter(reshape(SIGMA1WET{MONi,1},[],1),reshape(RAWSIGMA1WET{MONi,1},[],1),21,'b.')
31     scatter(reshape(SIGMA2WET{MONi,1},[],1),reshape(RAWSIGMA2WET{MONi,1},[],1),21,'k.')
32 end
33 plot([-1,1],[-1,1]);
34 set(gca,'XTick',[-0.3:0.1:1]); set(gca,'YTick',[-0.3:0.1:1]); set(gca,'FontSize',20)
35 xlabel(['The elements of \Sigma_{all} after adjust']); ylabel(['The elements of \Sigma_{all}↵
   before adjust'])
36
37
```

```matlab
1  %%%% A5-Simulation %%%%
2  clear; clc; close all;
3  load('PA4-SIGMA0WET(ADJ ESTIMATION).mat');
4  load('PA4-SIGMA1WET(ADJ ESTIMATION).mat');
5  load('PA4-SIGMA2WET(ADJ ESTIMATION).mat');
6  load('PA2-CVEC.mat');
7  DATEnum=datenum('Jan-01-1961'):datenum('Dec-31-2001'); TEMPDATEVEC=datevec(DATEnum);
8  DATE=[DATEnum',TEMPDATEVEC(:,1:3)];
9  SIMTIME=1000;
10 SIMWETcal=cell(SIMTIME,1);
11 SIMWETval=cell(SIMTIME,1);
12 tic
13 for SIMi=1:SIMTIME;
14     TEMPSIMWETTABLE=[DATE,NaN*ones(length(DATEnum),10)];
15     TEMPSIMZTABLE=[DATE,NaN*ones(length(DATEnum),10)];
16     SIGMAWETall=[SIGMA0WET{1,1}  ,SIGMA1WET{1,1}  ,SIGMA2WET{1,1}   ;
17                  SIGMA1WET{1,1}' ,SIGMA0WET{1,1}  ,SIGMA1WET{1,1}   ;
18                  SIGMA2WET{1,1}' ,SIGMA1WET{1,1}' ,SIGMA0WET{1,1}];
19     TEMPZSIM=reshape(mvnrnd(zeros(30,1),SIGMAWETall,1),10,3)';
20
21     TEMPSIMZTABLE(1,5:end)=TEMPZSIM(3,:);
22     TEMPSIMZTABLE(2,5:end)=TEMPZSIM(2,:);
23     TEMPSIMZTABLE(3,5:end)=TEMPZSIM(1,:);
24
25     TEMPSIMWETTABLE(1,5:end)=(TEMPZSIM(3,:)>(CVEC(:,1)'));
26     TEMPSIMWETTABLE(2,5:end)=(TEMPZSIM(2,:)>(CVEC(:,1)'));
27     TEMPSIMWETTABLE(3,5:end)=(TEMPZSIM(1,:)>(CVEC(:,1)'));
28
29     for DAYi=4:length(DATEnum);
30         TEMPMONi=TEMPSIMWETTABLE(DAYi,3);
31         CONDMEANCOEF=[SIGMA1WET{TEMPMONi,1},SIGMA2WET{TEMPMONi,1}]/([SIGMA0WET{TEMPMONi,1}, ↵
SIGMA1WET{TEMPMONi,1}',SIGMA1WET{TEMPMONi,1}',SIGMA0WET{TEMPMONi,1}]);
32         CONDCOVmatrix=SIGMA0WET{TEMPMONi,1}-CONDMEANCOEF*([SIGMA1WET{TEMPMONi,1},SIGMA2WET↵
{TEMPMONi,1}]');
33         TEMPSIMZTABLE(DAYi,5:end)=mvnrnd(CONDMEANCOEF*[TEMPSIMZTABLE(DAYi-1,5:end)';TEMPSIMZTABLE↵
(DAYi-2,5:end)'],CONDCOVmatrix,1)';
34         TEMPSIMWETTABLE(DAYi,5:end)=TEMPSIMZTABLE(DAYi,5:end)>(CVEC(:,TEMPMONi)');
35     end;
36     SIMWETcal{SIMi,1}=TEMPSIMWETTABLE(TEMPSIMWETTABLE(:,2)<=1985,:);
37     SIMWETval{SIMi,1}=TEMPSIMWETTABLE(TEMPSIMWETTABLE(:,2)>=1986,:);
38     fprintf(['SIMi=',num2str(SIMi),' TIME=',num2str(toc),'\n'])
39 end;
40 save('PA5-SIMWETcal.mat','SIMWETcal');
41 save('PA5-SIMWETval.mat','SIMWETval');
42
43
```

```matlab
function [ADJSIGMAWETall]=EIGCORRECTION(RAWSIGMAWETall,ERROR1,ERROR2)
[EIGVEC,EIGVAL]=eig(RAWSIGMAWETall+ERROR1*(1-eye(length(diag(RAWSIGMAWETall)))));
EIGVAL=diag(EIGVAL); EIGVAL(EIGVAL<ERROR2)=ERROR2;
ADJSIGMAWETall=diag(1./sqrt(diag(EIGVEC*diag(EIGVAL)*EIGVEC')))*(EIGVEC*diag(EIGVAL)*EIGVEC')*diag↙
(1./sqrt(diag(EIGVEC*diag(EIGVAL)*EIGVEC')));
```

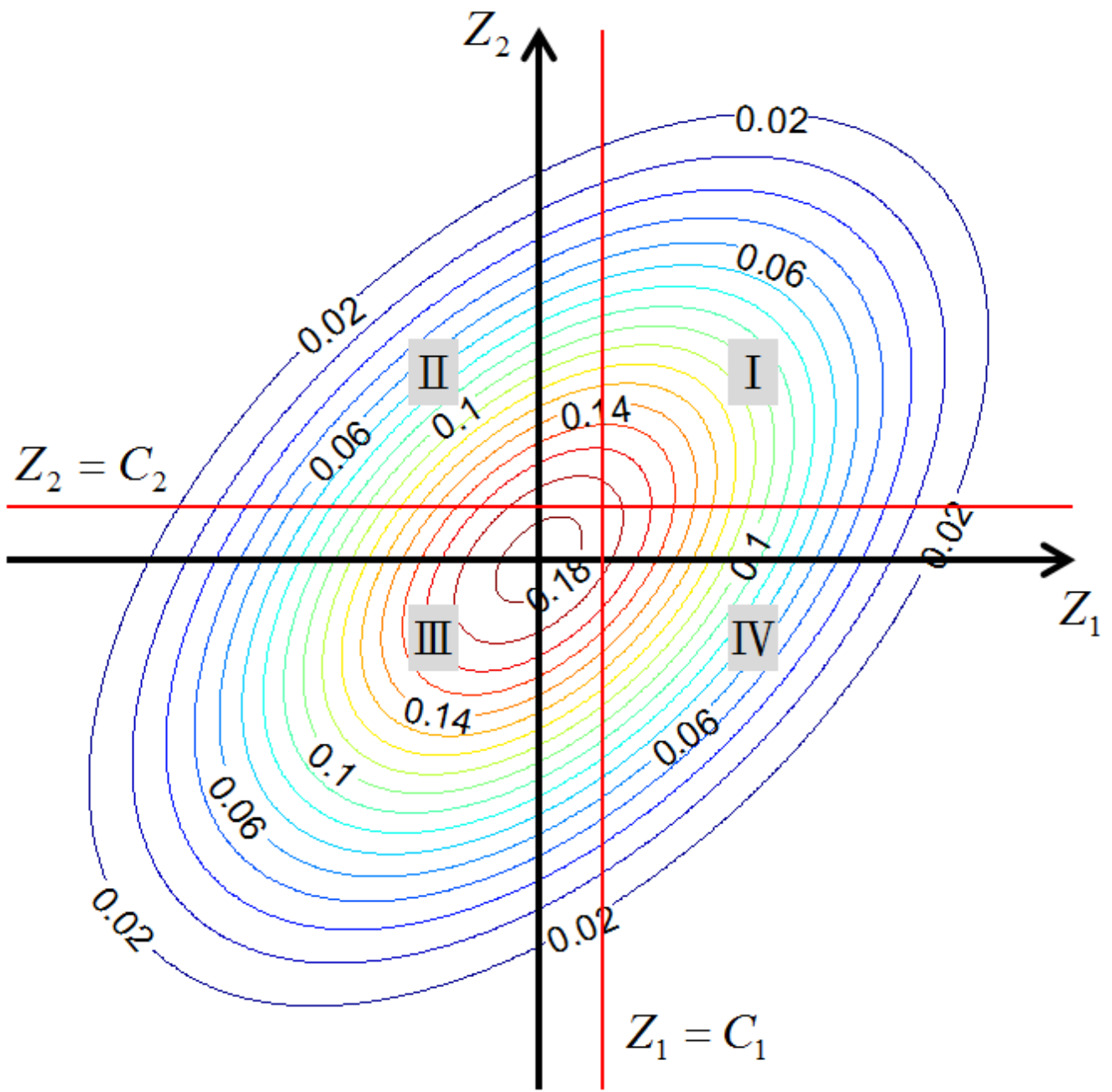

Figure 1. An example for illustrating the modeling concept. The contour lines represent the bivariate Gaussian density of $(Z_1, Z_2)^T$.

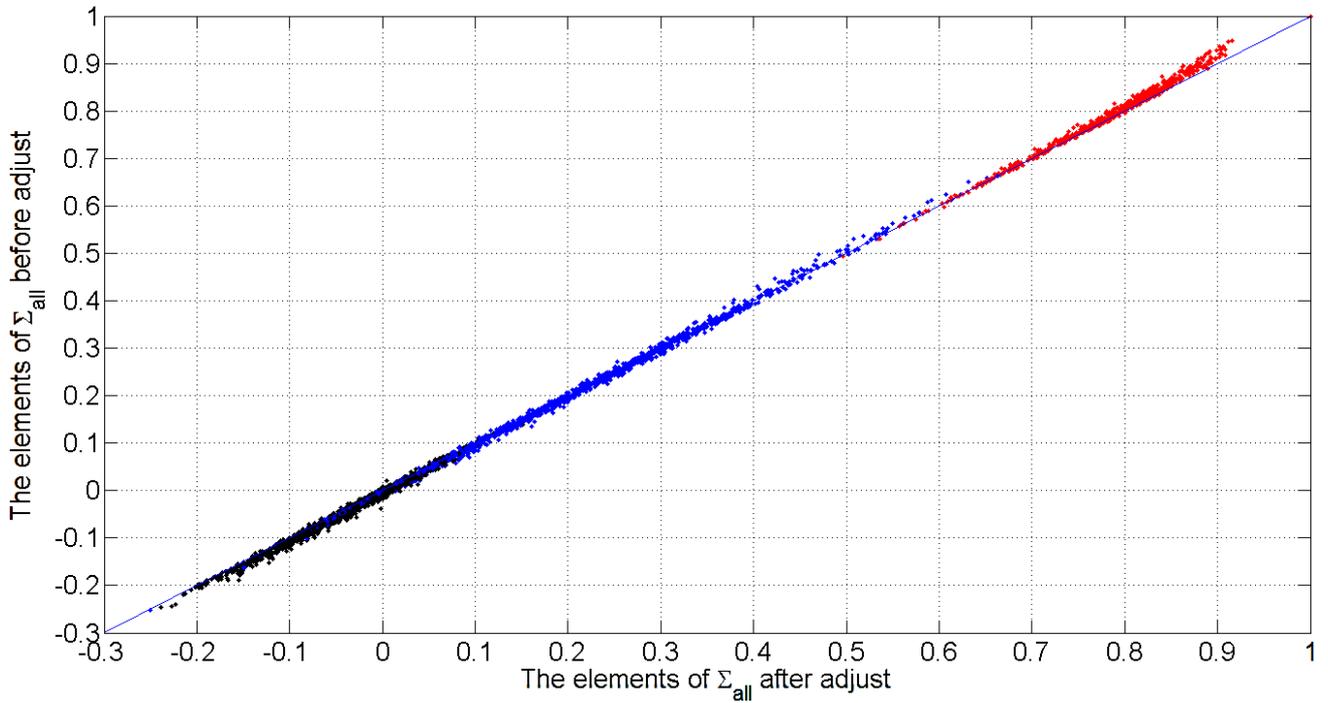

Figure 2. The elements of $\Sigma_{all}$ versus the elements of $\Sigma_{all}^{adj.}$ plot. The twelve estimated $\Sigma_{all}$ matrices were obtained from modeling the daily precipitation occurrences of the twelve calendar months of the study area with the lag-2 model, respectively. $\varepsilon_2$ is set to be 0.05 for adjusting the $\Sigma_{all}$ matrices. Red, blue, and black dots correspond to interstation, lag-1 interstation, and lag-2 interstation correlations, respectively.

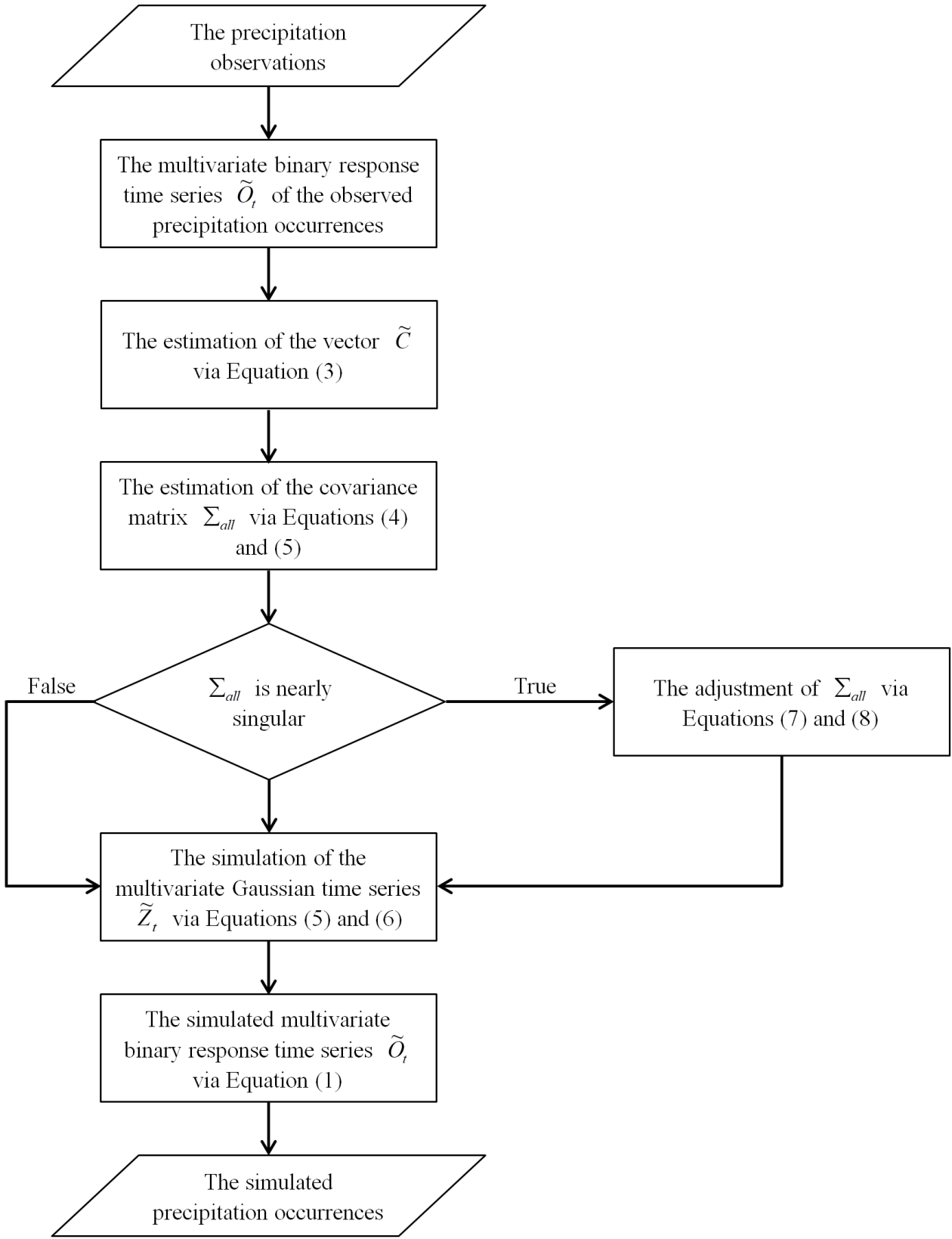

Figure 3. The flowchart that describes the modeling and the simulation procedures of the application of the proposed method to precipitation occurrences.

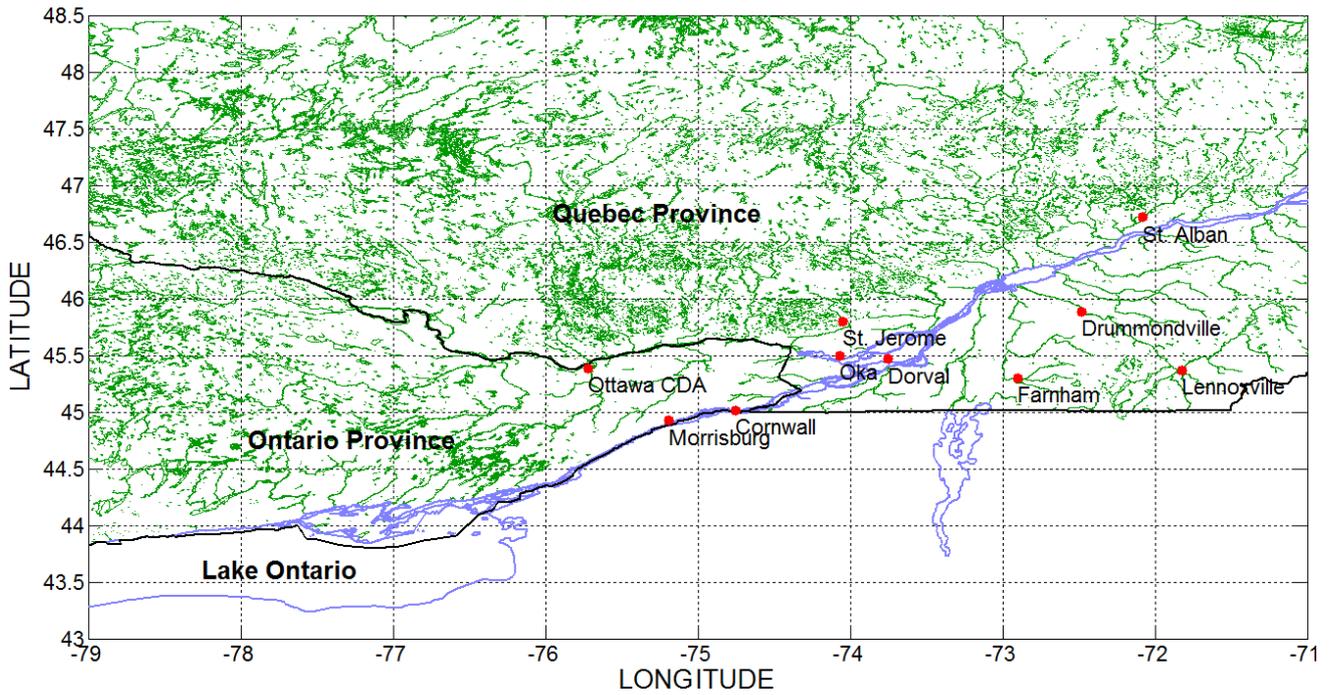

Figure 4. The study area and the distribution of the raingauge sites. Green lines are polygons of lakes and rivers, blue lines are polygons of coastal waters, and black lines represent the borders of the provinces (the vector data used to prepare this map comes from Statistics Canada).

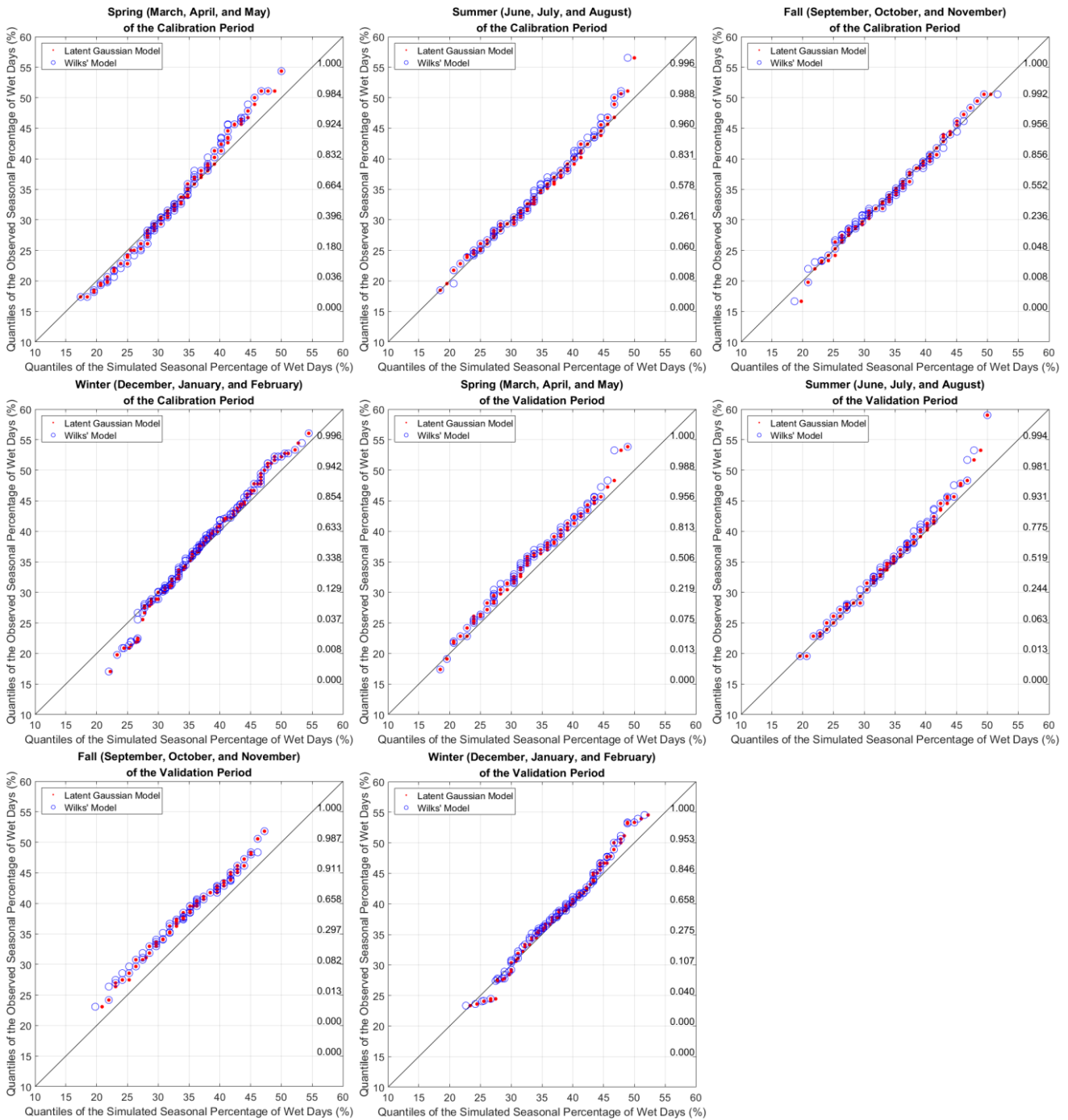

Figure 5. The quantile to quantile comparison between the observed and simulated percentages of wet days at the seasonal scale. The horizontal axis represents the median of the corresponding simulated quantiles. The corresponding cumulative probabilities of the observed seasonal percentages of wet days are labeled on the right-hand side of each plot.

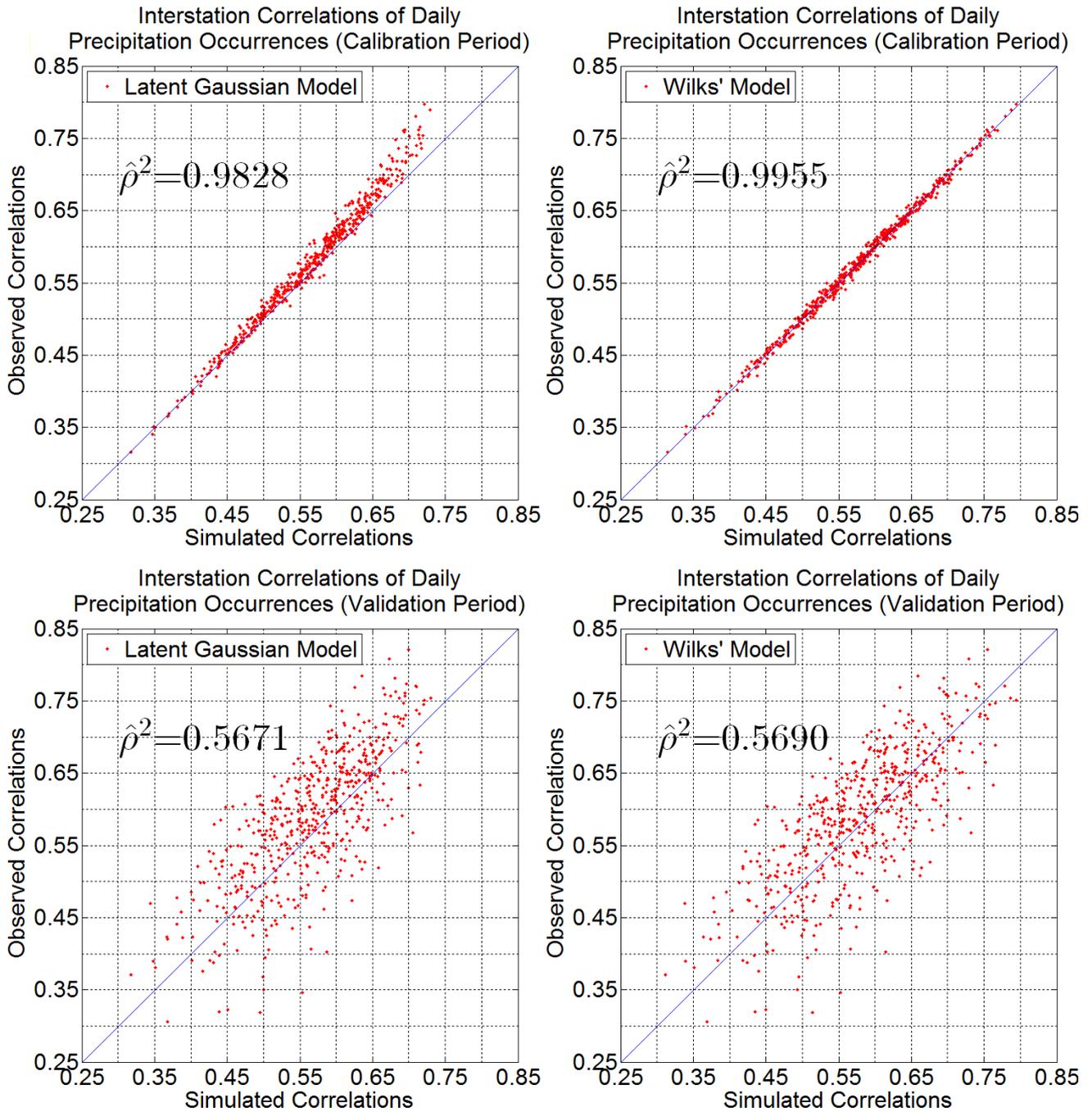

Figure 6. The scatter plots of the observed and simulated pairs of the interstation correlations of the daily precipitation occurrences. The interstation correlations are calculated from all twelve calendar months and all station pairs. The top and bottom are the comparison results in the calibration and validation periods, respectively. The left and right are the comparison results of the latent Gaussian model and Wilks' model, respectively.

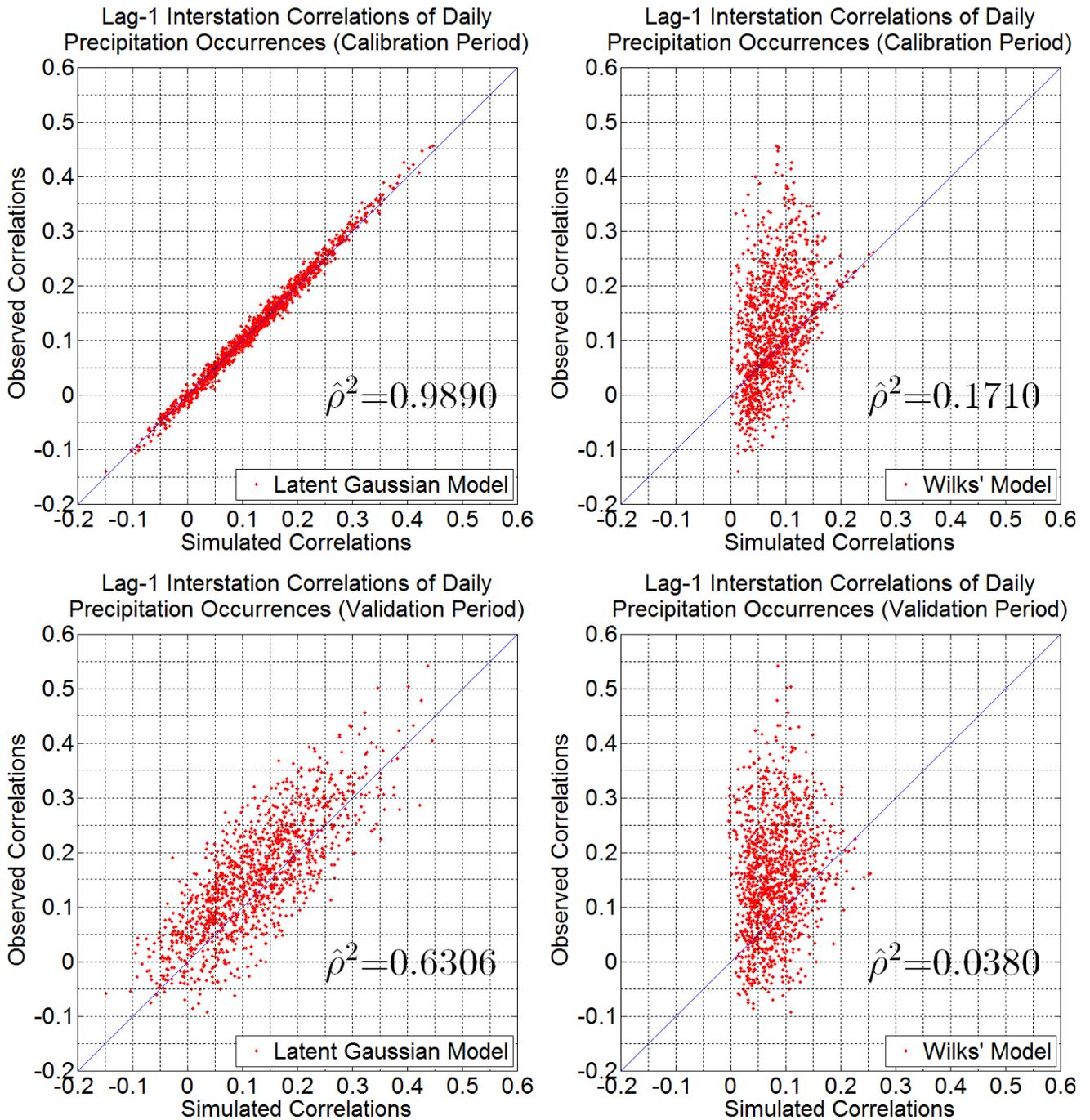

Figure 7. The scatter plots of the observed and simulated pairs of the lag-1 interstation correlations of the daily precipitation occurrences. The interstation correlations are calculated from all twelve calendar months and all station pairs. The top and bottom are the comparison results in the calibration and validation periods, respectively. The left and right are the comparison results of the latent Gaussian model and Wilks' model, respectively.

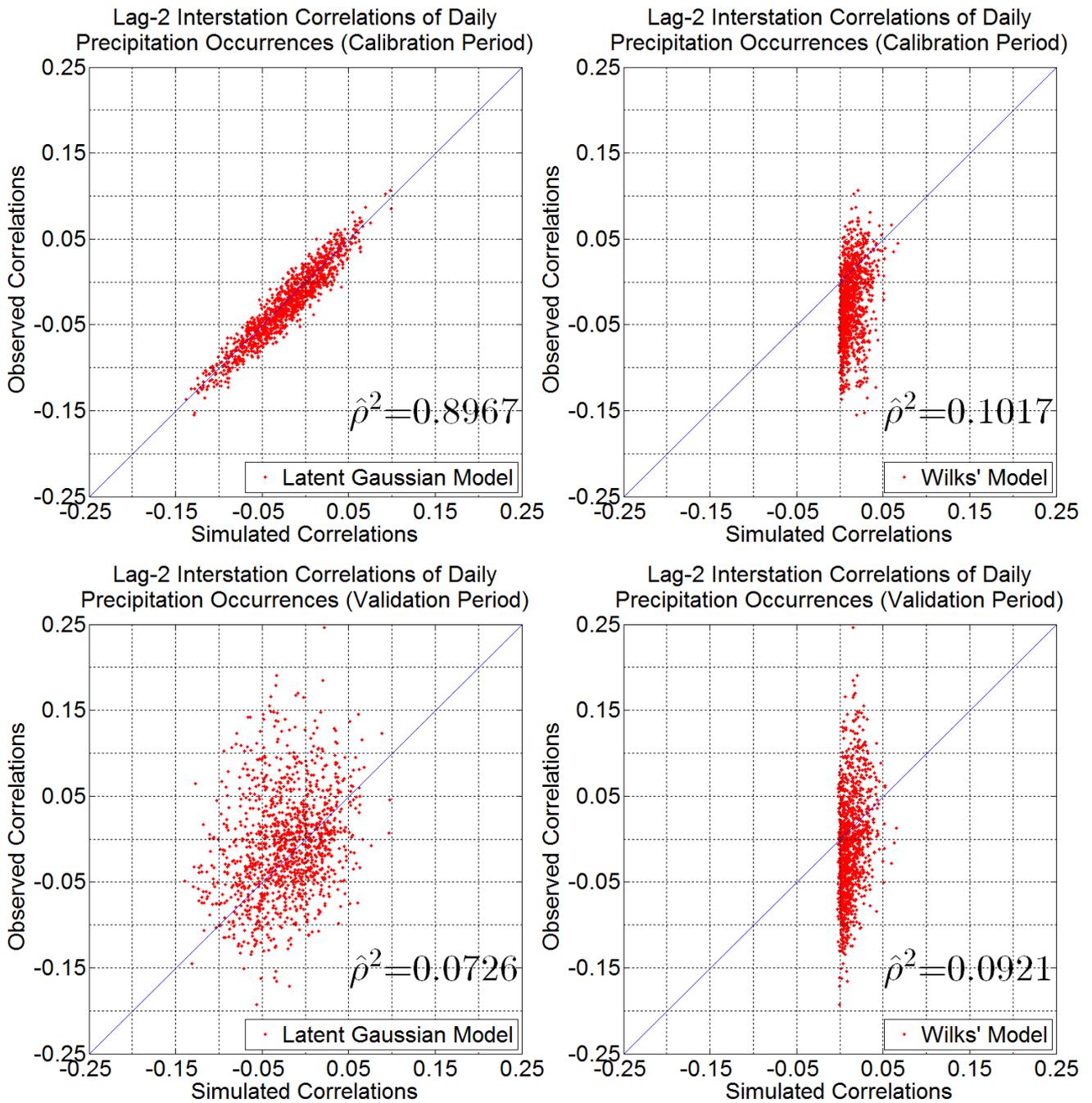

Figure 8. The scatter plots of the observed and simulated pairs of the lag-2 interstation correlations of the daily precipitation occurrences. The interstation correlations are calculated from all twelve calendar months and all station pairs. The top and bottom are the comparison results in the calibration and validation periods, respectively. The left and right are the comparison results of the latent Gaussian model and Wilks' model, respectively.

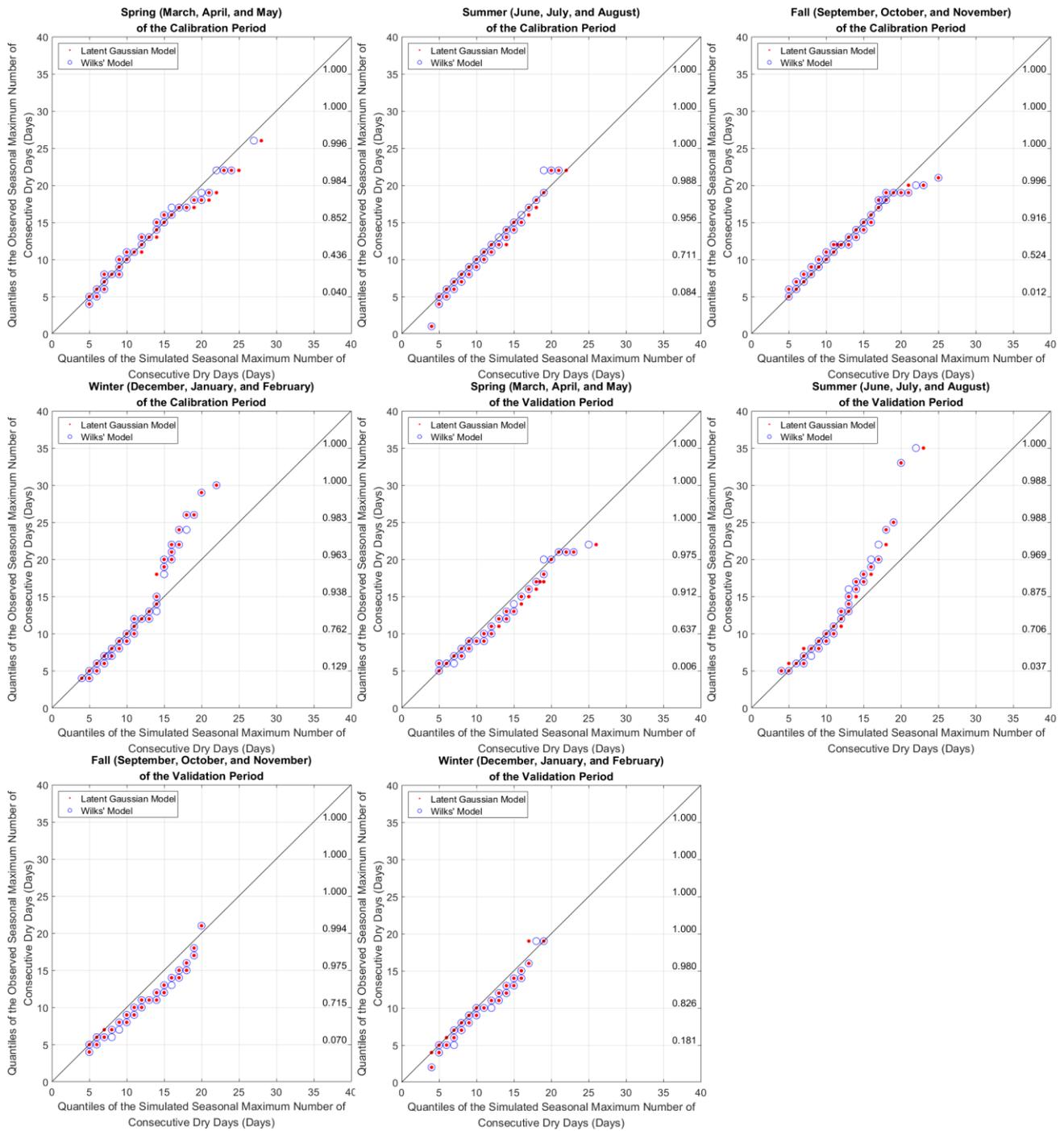

Figure 9. The quantile to quantile comparison between the observed and simulated maximum numbers of consecutive dry days at the seasonal scale. The horizontal axis represents the median of the corresponding simulated quantiles. The corresponding cumulative probabilities of the observed seasonal maximum numbers of consecutive dry days are labeled on the right-hand side of each plot.

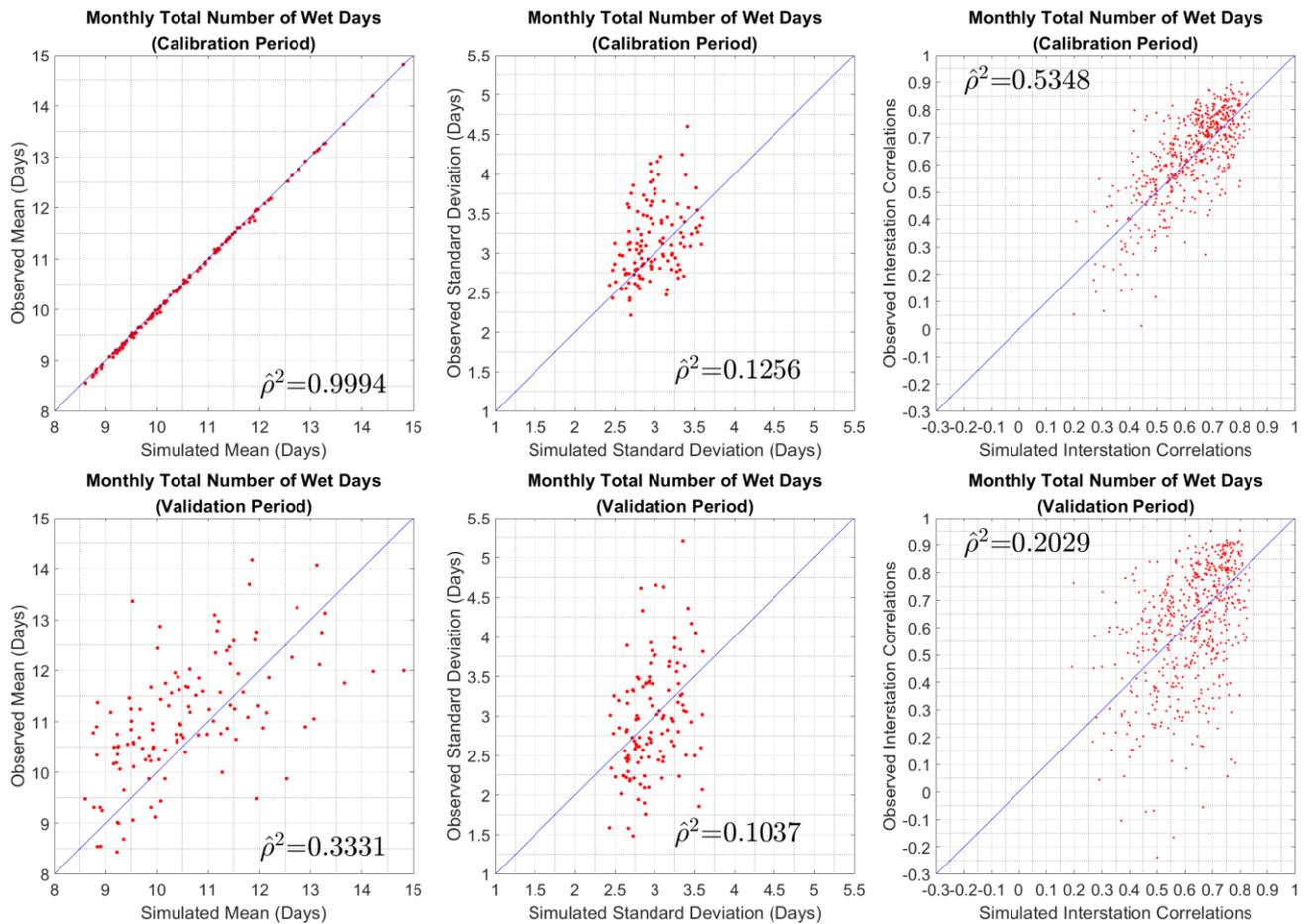

Figure 10. The scatter plots of the observed and simulated pairs of the mean, the standard deviation, and the interstation correlation of the monthly total number of wet days. The means and the standard deviations are calculated from all twelve calendar months and all the stations while the interstation correlations are calculated from all twelve calendar months and all the station pairs. The top and bottom are the comparison results in the calibration and the validation periods, respectively. Due to the missing observations of the precipitation record, the monthly total number of wet days of a month is the mean wet day per day of that month times the number of days of that month.

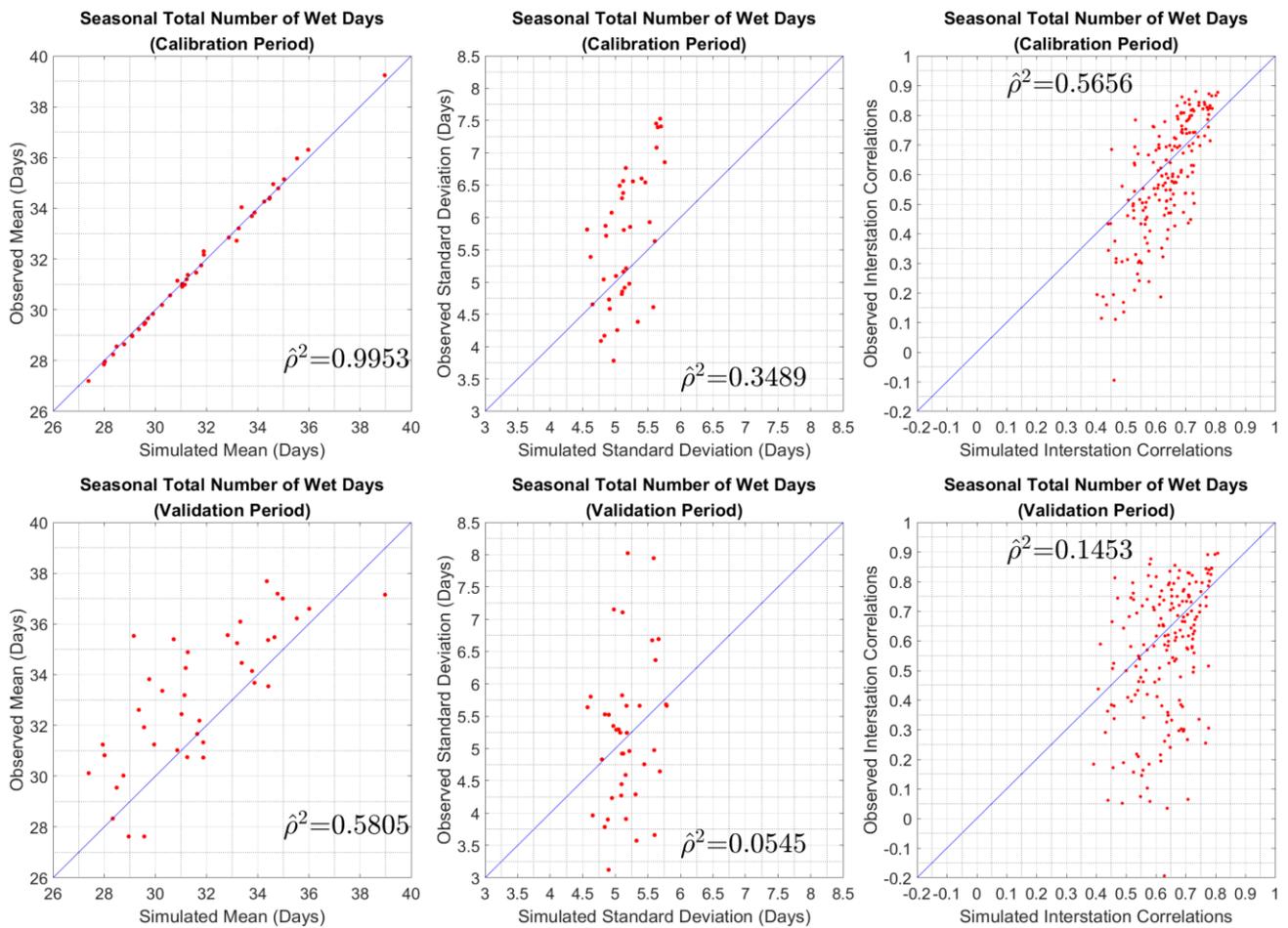

Figure 11. The scatter plots of the observed and simulated pairs of the mean, the standard deviation, and the interstation correlation of the seasonal total number of wet days. The means and the standard deviations are calculated from all four seasons and all the stations while the interstation correlations are calculated from all four seasons and all the station pairs. The top and bottom are the comparison results in the calibration and the validation periods, respectively. Due to the missing observations of the precipitation record, the seasonal total number of wet days of a season is the mean wet day per day of that season times the number of days of that season.

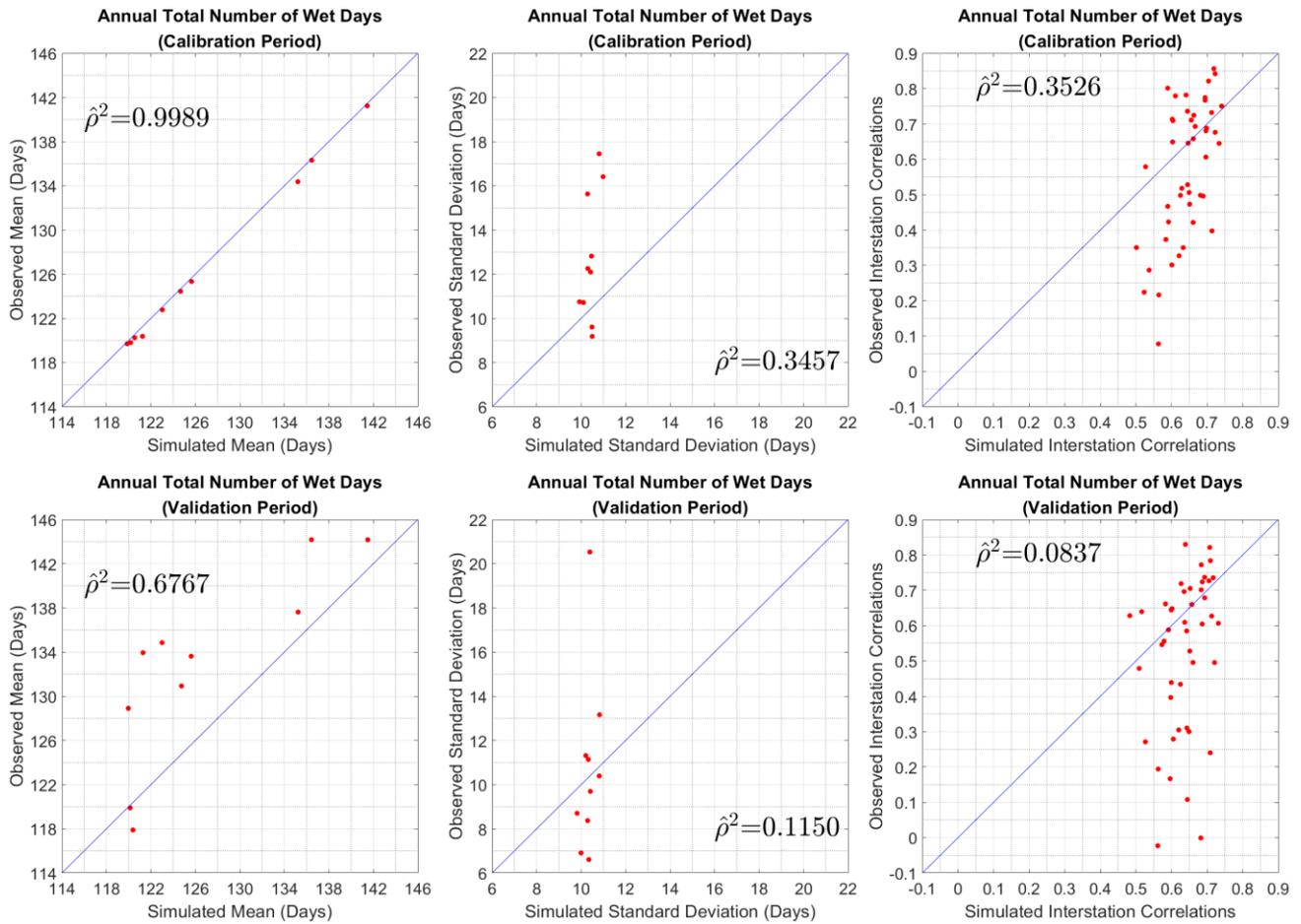

Figure 12. The scatter plots of the observed and simulated pairs of the mean, the standard deviation, and the interstation correlation of the annual total number of wet days. The means and the standard deviations are calculated from all the stations while the interstation correlations are calculated from all the station pairs. The top and bottom are the comparison results in the calibration and the validation periods, respectively. Due to the missing observations of the precipitation record, the annual total number of wet days of a year is the mean wet day per day of that year times the number of days of that year.